\documentclass[conference]{IEEEtran}
\usepackage[absolute,showboxes,overlay]{textpos}
\usepackage{hyperref}

\TPMargin{5pt}

\newcommand{\copyrightstatement}{
    \begin{textblock}{0.84}(0.08,0.93)    
         \noindent
         \footnotesize
         \copyright 2021 IEEE. Personal use of this material is permitted. Permission from IEEE must be obtained for all other uses, in any current or future media, including reprinting/republishing this material for advertising or promotional purposes, creating new collective works, for resale or redistribution to servers or lists, or reuse of any copyrighted component of this work in other works. {Cite from IEEE: }\href{<https://ieeexplore.ieee.org/document/9685255>}{DOI No. 10.1109/GLOBECOM46510.2021.9685255}
    \end{textblock}
}

\IEEEoverridecommandlockouts
\usepackage{cite}
\usepackage{amsmath,amssymb,amsfonts}
\usepackage{graphicx}
\usepackage{textcomp}
\usepackage{xcolor}

\usepackage{subfigure}
\usepackage{fancyhdr}
\usepackage{algpseudocode}
\usepackage{booktabs} 
\usepackage{multirow}
\usepackage{multicol}
\usepackage{setspace}

\def\BibTeX{{\rm B\kern-.05em{\sc i\kern-.025em b}\kern-.08em
    T\kern-.1667em\lower.7ex\hbox{E}\kern-.125emX}}
\begin{document}
\copyrightstatement
\title{Deep Generative Model for Simultaneous Range Error Mitigation and Environment Identification}

\author{
\IEEEauthorblockN{
Yuxiao~Li\IEEEauthorrefmark{1},
Santiago~Mazuelas\IEEEauthorrefmark{2}, and
Yuan~Shen\IEEEauthorrefmark{1}}
\IEEEauthorblockA{\IEEEauthorrefmark{1}
Department of Electronic Engineering,
Tsinghua University,
Beijing, China \\
}
\IEEEauthorblockA{\IEEEauthorrefmark{2}
BCAM-Basque Center for Applied Mathematics, and IKERBASQUE-Basque Foundation for Science, Bilbao, Spain \\
Email: li-yx18@mails.tsinghua.edu.cn,
smazuelas@bcamath.org,
shenyuan\_ee@tsinghua.edu.cn
}}

\maketitle

\begin{abstract}

Received waveforms contain rich information for both range information and environment semantics. However, its full potential is hard to exploit under multipath and non-line-of-sight conditions. This paper proposes a deep generative model (DGM) for simultaneous range error mitigation and environment identification. In particular, we present a Bayesian model for the generative process of the received waveform composed by latent variables for both range-related features and environment semantics. The simultaneous range error mitigation and environment identification is interpreted as an inference problem based on the DGM, and implemented in a unique end-to-end learning scheme. Comprehensive experiments on a general Ultra-wideband dataset demonstrate the superior performance on range error mitigation, scalability to different environments, and novel capability on simultaneous environment identification.

\end{abstract}

\begin{IEEEkeywords}
Deep Generative Model, Range Error Mitigation, Environment Identification, Non-Linear Signal Processing, Bayesian Model
\end{IEEEkeywords}

\section{Introduction}
\label{sec:intro}

As a fundamental description of the channel, the received waveform represents how a signal propagates from the transmitter to the receiver in a multipath channel. Received waveforms inherently contain information both range-related and environment-related, and have been widely exploited in many localization algorithms for harsh environments \cite{MazConAllWin:J18}. Exploiting semantic features from received waveforms can provide enhanced localization, and is critical for beyond fifth-generation (B5G) network requirements \cite{YuLiuMeyConWin:J20}.

Typical localization systems obtain received signals corresponding with specific radio-frequency technologies including Wi-Fi \cite{KotJosBha:C15}, mmWave \cite{MenMeyBauWin:J19}, and ultra-wideband (UWB) \cite{WymMarGifWin:J12}. Among these technologies, UWB transmission is the most promising with a bandwidth over $500$ MHz and extremely short transmitted pulses, which allow for a finer time resolution of multipath signals.
\begin{figure}[htbp]
    \begin{center}
        \includegraphics[width=0.45\textwidth]{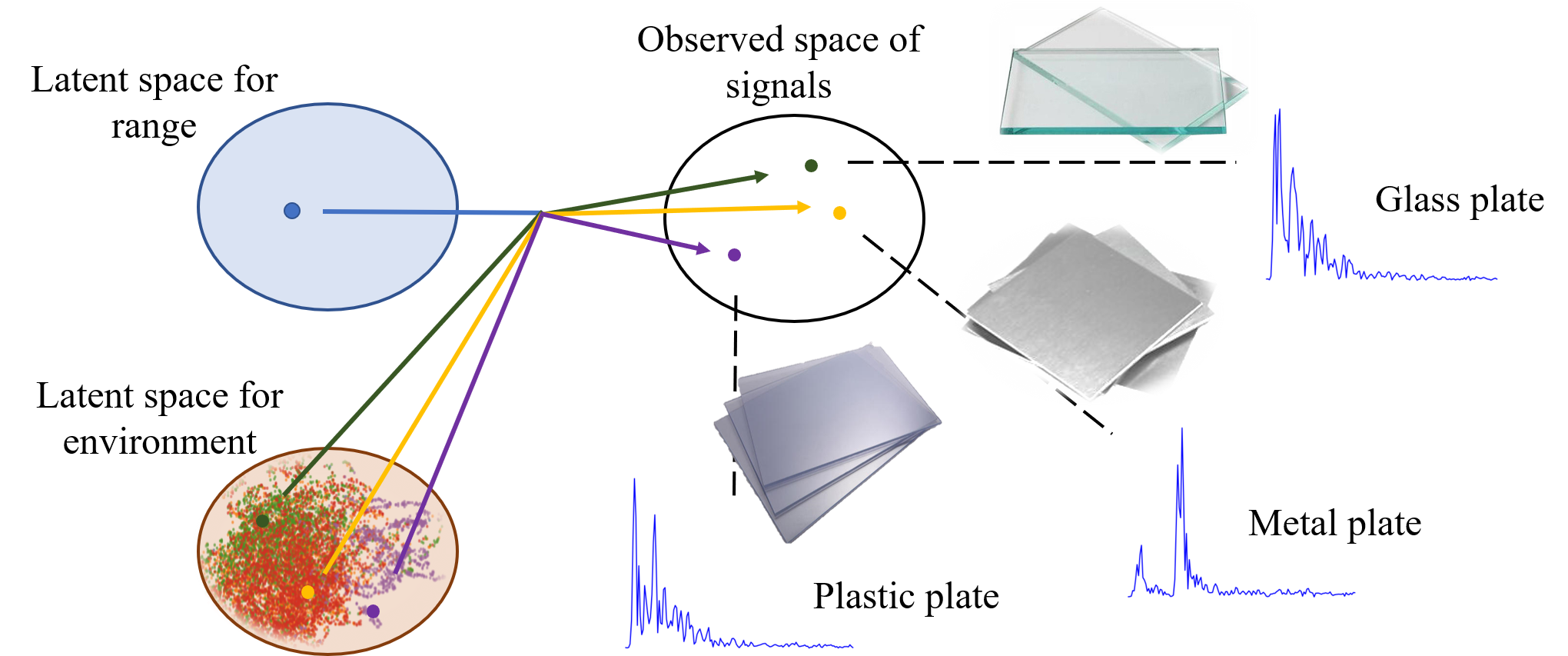}
        \caption{Illustrations of the proposed generative model for received waveforms. Waveform samples in data space are generated with latent variables for range-related features and environment semantics. Different environment semantics lead to different output waveforms.}
        \label{fig:graphic}
    \end{center}
\end{figure}
However, the capability of these radio-frequency systems in practical deployment is still limited and hurdled by a number of technical challenges, especially in harsh environments. These devices often adopt the first estimated delay as the line-of-sight (LOS) path to perform range estimation \cite{KotJosBha:C15}. Though easy to implement, such method tend to introduce a positive bias in range estimation caused by the non-line-of-sight (NLOS) propagation, as well as a cluttering noise due to the multi-path effect \cite{SheWymWin:J10}. Moreover, the usage of received waveforms to identify different environments currently provides a coarse LOS or NLOS classification, leaving out richer environmental semantics like geometric layouts of the room and materials of the blocking objects \cite{WymMarGifWin:J12}. Therefore, techniques to improve range error mitigation and obtain detailed environment identification are imperative for wireless networks \cite{KanRap:J21}.

Existing methods address separately range error mitigation and environment identification. For range error mitigation, conventional methods detect NLOS propagation and assign different weights to LOS and NLOS range estimations for positional purpose \cite{KhaKarMou:J10}. Early machine learning methods, such as SVM \cite{WymMarGifWin:J12}, use hand-crafted features \cite{MazConAllWin:J18} to represent received signals and learn the range error from a regression problem. Recently, several deep learning methods have been proposed where the whole waveform is utilized as input to learn a regressor of the range error \cite{AngMazSalFanChi:J20}. Though enjoying a significant improvement in performance using full waveforms, such deep learning methods suffer from generalization problems and are prone to overfitting.

Existing methods for environment identification aim to determine coarse LOS or NLOS conditions instead of more detailed environment semantics. These methods use characteristic features of waveform to determine the LOS or NLOS conditions. As in range error mitigation, these features are either hand-crafted conducted by SVM \cite{XiaWenMarTriBluFro:J15}, relevance vector machine (RVM) \cite{NguJeoShiWin:J15}, or data-driven conducted by neural networks \cite{DecOrdFerHe:J18}.

We propose a deep generative model (DGM) for simultaneous range error mitigation and environment identification, with a generative modeling illustrated in Fig.\ref{fig:graphic}. 
Specifically, we present a Bayesian model for the generative process of the received waveform, where the range-related features and environmental semantics are disentangled from waveforms via a modified variational auto-encoder (VAE). Simultaneous inference is carried out based on these features by two sub neural modules. The presented method is technology-agnostic and is applicable to any technology providing received waveforms.



The remaining sections are organized as follows. 
Section \ref{sec:model} introduces the proposed DGM, including the Bayesian modeling and network learning scheme employed. The performance of the proposed method on both tasks is evaluated with a case study in Section \ref{sec:exp}. Finally, Section \ref{sec:con} concludes the paper.

\section{Deep Generative Model}
\label{sec:model}

In this section we propose a deep learning method for simultaneous range error mitigation and environment identification. We first describe the problem of exploiting received waveforms to obtain richer semantics. Then a Bayesian model is presented involving a range-related variable and an environment-related variable. Afterwards, we introduce the network implementation and learning scheme of the proposed method, illustrated in Fig.\ref{fig:str}.

\subsection{Problem Statement}
\label{sec:model_pro}

Given the transmitted waveform $\mathbf{s}$, the received signal $\mathbf{x}$ can be expressed as follows,
\begin{equation}  \label{eq:observe}
    \mathbf{x}(t) = \sum_{l} \alpha_l \mathbf{s}(t-\tau_l) + \mathbf{n}(t), ~t\in [0, T_{ob}]
\end{equation}
\noindent where $\mathbf{s}(t)$ denotes the transmitted waveform, $L$, denotes the number of multi-path components, $\alpha_l$, and $\tau_l$ are the amplitude and propagation delay of the $l$th component, respectively, $\mathbf{n}(t)$ represents an additive white Gaussian noise (AWGN), and $[0, T_{ob}]$ is the observation interval.

Suppose different environment scenarios are annotated by labels with discrete values, denoted as $k$ in $\{1, 2, \ldots, K\}$. Techniques for environment identification utilizes the whole waveform $\mathbf{x}$ to estimate scenario $k$, while range error mitigation estimate the non-negative bias $\Delta d$ between the true distance $d$ and the measured distance $d_M$ from the delay components. Both tasks involve complex semantics caused by different environments.
It is difficult to capture all of the environment factors in one theoretic model. Nevertheless, some observations from the received waveforms can be concluded to give helpful insights \cite{WymMarGifWin:J12,MazConAllWin:J18}: 
i) received waveforms show different characteristics in different environments; ii) the distinction between LOS and NLOS conditions is rather coarse to exploit the inherent semantics of waveforms. 

This paper develops a more general approach to exploit the inherent semantics of received waveforms, which can disentangle the related features in received waveforms and simultaneously conduct range error mitigation and environment identification.

\subsection{Bayesian Model for Received Waveform}

  \begin{figure*}[htbp]
      \centerline{
      \includegraphics[width=1.0\textwidth]{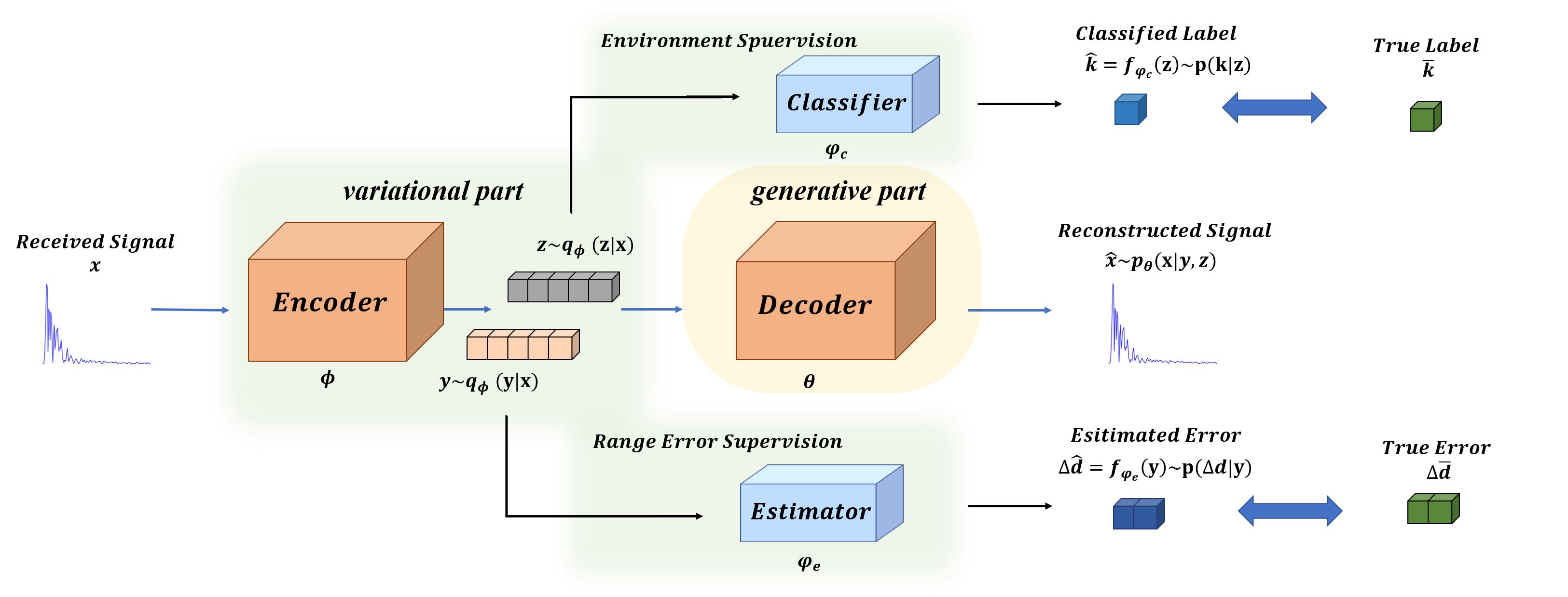}
      }
      \caption{Network structure of the proposed DGM for simultaneous range error mitigation and environment identification. Waveform samples are encoded into two latent variables with the guidance of real environment labels and range errors from dataset. The estimations are obtained via sub modules, with the according features extracted by the encoder from waveform as inputs.}
      \label{fig:str}
  \end{figure*}

We model the generative process of the waveform $\mathbf{x}$ involving two independent latent variables: $\mathbf{y}$ for range-related features and $\mathbf{z}$ for environment-related features. The generation from the latent space to the data space can be obtained via the likelihood distribution $p(\mathbf{x}|\mathbf{y},\mathbf{z})$. 
The generative process consists of three sequential steps:

\begin{enumerate}
    \item A value $\textnormal{y}^{(i)}$ is generated from distribution $p(\mathbf{y})$ of the latent variable for range-related feature.
    \item A value $\textnormal{z}^{(i)}$ is generated from distribution $p(\mathbf{z})$ of the latent variable for environment-related feature.
    \item A value $\textnormal{x}^{(i)}$ for a UWB measurement sample is generated from the conditional distribution $p(\mathbf{x}|\mathbf{y}^{(i)}, \mathbf{z}^{(i)})$.
\end{enumerate}

Therefore, the generation of a measurement sample can be obtained by sampling the distribution on the two latent variables, and the distribution of data conditioned on these samples, i.e.,
\begin{equation}  \label{qe:generation}
    \textnormal{x}^{(i)} = g(\textnormal{y}^{(i)}, \textnormal{z}^{(i)}) \sim p(\mathbf{x}|\textnormal{y}^{(i)},\textnormal{z}^{(i)})
\end{equation}

\noindent where $\textnormal{y}^{(i)}\in\mathbb{R}^{D_r}$ and $\textnormal{z}^{(i)}\in\mathbb{R}^{D_e}$. 

The estimations of range error and environment label from waveform $\mathbf{x}^{(i)}$ can be obtained from the conditional distributions $p(\Delta d|\mathbf{x}^{(i)})$ and $p(k|\mathbf{x}^{(i)})$. With the two latent variables defined above, the problem can be transferred to estimating $p(\Delta d|\mathbf{y}^{(i)})$ and $p(k|\mathbf{z}^{(i)})$ instead, with $\mathbf{y}^{(i)}, \mathbf{z}^{(i)}$ being the latent variables corresponding with $\mathbf{x}^{(i)}$.



\subsection{Network Learning Scheme}

Given the observed measurements, the estimated range error $\Delta \hat{d}$ is obtained by the range-related variable $\mathbf{y}$, and the predicted environment label $\hat{k}$ is obtained by the environment-related variable $\mathbf{z}$. Both variables are obtained from the waveform $\mathbf{x}$ using the deep learning method described in the following.

Given a dataset $\mathcal{D}=\{\mathbf{x}^{(i)}, \Delta {d}^{(i)}, {k}^{(i)}\}_{i=1}^N$ consisting of $N$ i.i.d. samples with paired waveform $\mathbf{x}$, range error $\Delta {d}$, and environment label ${k}$. The network structure consists of three sub neural modules: an auto-encoder (AE) for the disentanglement of latent variables,
an estimator to predict range error,
and a classifier to predict environment label.
The whole network structure is illustrated in Fig.\ref{fig:str}.
Specifically, the training phase consists of simultaneous-learning of the three neural modules:

\begin{enumerate}
    \item the AE is learned on waveform sample $\mathbf{x}^{(i)}\sim\mathcal{D}$ and disentangles two latent variables $\mathbf{y}^{(i)}$ and $\mathbf{z}^{(i)}$ in the bottleneck;
    \item the estimator is learned on $\mathbf{y}^{(i)}$ to obtain the range error $\Delta \hat{d}^{(i)}$ with the supervision of $\Delta d^{(i)}$ in the dataset;
    \item the classifier is learned on $\mathbf{z}^{(i)}$ to obtain the environment label $\hat{k}^{(i)}$ with the supervision of $k^{(i)}$ in the dataset.  
\end{enumerate}

In the testing phase, network parameters are learned and frozen. Given any received waveform $\mathbf{x}^{(j)}$, simultaneous error mitigation and environment identification are conducted on pure waveform data with the following two steps:

\begin{enumerate}
    \item feed the waveform sample $\mathbf{x}^{(j)}$ into the encoder of VAE and get range code $\mathbf{y}^{(j)}$ and environment code $\mathbf{z}^{(j)}$;
    \item feed code $\mathbf{y}^{(j)}$ into the estimator and get estimated range error $\Delta \hat{d}^{(j)}$;
    \item feed code $\mathbf{z}^{(j)}$ into the classifier and get environment label $\hat{k}^{(j)}$.
\end{enumerate}

Since the whole network is learned in a unified scheme, range error mitigation and environment identification can be conducted simultaneously in the testing step.

\begin{table*}[t]
\caption{Quantitative results on range error mitigation and environment identification with five different environment settings.}
\label{tab:exp_comp}
\begin{center}
\begin{small}
\begin{sc}
\begin{tabular}{lccccccc}
\toprule
\multirow{3}{*}{\shortstack{Environment \\Scenarios}} & \multicolumn{5}{c}{Range Error Mitigation} & \multicolumn{2}{c}{Environment Identification}  \\
& {Unmitigated} & \multicolumn{2}{c}{SVR} & \multicolumn{2}{c}{DGM} & SVC & DGM \\
& MAE & RMSE & MAE & RMSE & MAE & Accuracy & Accuracy   \\
\midrule
\textit{Room Full}     &0.1084 &0.1553 &0.0895 &0.0568 &0.0163 &0.4859 &0.6203  \\
\textit{Room Rough}  &0.1066 &0.1535 &0.0886  &0.0539 &0.0157 &0.9945 &0.9998  \\
\textit{Room Part}  &0.1118 &0.1677 &0.0916 &0.0462 &0.0078 &0.7330 &0.8551  \\
\midrule
\textit{Obstacle Full}    &0.1271 &0.1746 &0.1018 &0.0678 &0.0185 &0.3129 &0.3334  \\
\textit{Obstacle Rough}  &0.1571 &0.2083 &0.1193 &0.0878 &0.0232 &0.8650 &0.9362  \\
\bottomrule
\end{tabular}
\end{sc}
\end{small}
\end{center}
\end{table*}


\subsection{Formulation of the Objective Function}

Suppose the auto-encoder (AE) is denoted as $(\operatorname{g}_{\text{enc}}, \operatorname{g}_{\text{dec}})$ 
with parameters $\{\boldsymbol{\phi},\boldsymbol{\theta}\}$, where $\operatorname{g}_{\text{enc}}(\cdot; \boldsymbol{\phi}):\mathbf{x}\to \mathbf{y}, \mathbf{z}$ and $\operatorname{g}_{\text{dec}}(\cdot; \boldsymbol{\theta}):\mathbf{y}, \mathbf{z}\to \mathbf{x}$.
The range error estimator is denoted as $\operatorname{f}_{\text{est}}$ with parameter $\boldsymbol{\varphi}_e$, and the environment classifier as $\operatorname{f}_{\text{cls}}$ with parameter $\boldsymbol{\varphi}_c$, where $\operatorname{f}_{\text{est}}(\cdot; \boldsymbol{\varphi}_e): \mathbf{y}\to d$ and $\operatorname{f}_{\text{cls}}(\cdot; \boldsymbol{\varphi}_c): \mathbf{z}\to k$. The objective function w.r.t. dataset $\mathcal{D}$ and parameters $\boldsymbol{\phi}$, $\boldsymbol{\theta}$ and $\boldsymbol{\varphi}$ is composed of three loss terms: \textit{i)} a reconstruction loss to regularize the outputs of AE, \textit{ii)} a regression loss to make $\Delta \hat{d}$ close to the real error $\Delta {d}$, and \textit{iii)} a classification loss to make $\hat{k}$ close to label ${k}$, i.e.,
\begin{equation}  \label{eq:objective}
    \begin{aligned}
       \mathbb{L}(\mathcal{D};\boldsymbol{\phi},\boldsymbol{\theta},\boldsymbol{\varphi}) = \mathbb{L}_{rec}\big(\hat{\mathbf{x}}, \mathbf{x}\big) + \mathbb{L}_{est}\big(\Delta \hat{d}, \Delta {d}\big) + \mathbb{L}_{cls}\big(\hat{k}, {k}\big)
    \end{aligned}
\end{equation}
\noindent where related variables from AE are $\mathbf{\hat{x}}=g_{\text{dec}}\big(g_{\text{enc}}(\mathbf{x};\boldsymbol{\phi}); \boldsymbol{\theta}\big)$, the variable from the estimator is $\Delta \hat{d}=f_{\text{est}}\big(\mathbf{y};{\boldsymbol{\varphi}_e}\big)$, and the variable from the classifier is $\hat{k}=f_{\text{cls}}\big(\mathbf{z};{\boldsymbol{\varphi}_c}\big)$.

The reconstruction loss term in equation~\eqref{eq:objective} is given by:
\begin{equation}  \label{eq:rec}
    \begin{aligned}
       \mathbb{L}_{rec}(\mathcal{D};\boldsymbol{\phi},\boldsymbol{\theta}) &= \sum_{i=1}^N \Vert \mathbf{x}^{(i)} - \hat{\mathbf{x}}^{(i)} \Vert_2^2  \\
       &= \sum_{i=1}^N \Vert \mathbf{x}^{(i)} - \operatorname{g}_{\text{dec}}\big(\operatorname{g}_{\text{enc}}(\mathbf{x}^{(i)};\boldsymbol{\phi});\boldsymbol{\theta}\big) \Vert_2^2
    \end{aligned}
\end{equation}

The loss terms for the estimator and the classifier in equation~\eqref{eq:objective} are given by:
\begin{equation}  \label{eq:est}
    \begin{aligned}
       \mathbb{L}_{est}(\mathcal{D};\boldsymbol{\theta},\boldsymbol{\varphi}_\text{e}) &= \sum_{i=1}^N \Vert {\Delta \hat{d}}^{(i)} - {\Delta {d}}^{(i)}\Vert^2  \\
       &= \sum_{i=1}^N \Vert \operatorname{f}_{\text{est}}(\operatorname{g}_{enc}({\mathbf{x}}^{(i)};\boldsymbol{\phi});\boldsymbol{\varphi}_e) - {\Delta {d}}^{(i)}\Vert^2
    \end{aligned}
\end{equation}
\begin{equation}  \label{eq:cls}
    \begin{aligned}
       \mathbb{L}_{cls}(\mathcal{D};\boldsymbol{\theta},\boldsymbol{\varphi}_\text{c}) &= \sum_{i=1}^N \Vert {\hat{k}}^{(i)}- {{k}}^{(i)}\Vert^2  \\
       &= \sum_{i=1}^N \Vert \operatorname{f}_{\text{cls}}(\operatorname{g}_{enc}({\mathbf{x}}^{(i)};\boldsymbol{\phi});\boldsymbol{\varphi}_c)- {{k}}^{(i)}\Vert^2
    \end{aligned}
\end{equation}

Implemented by the proposed DGM with loss function Eq.\eqref{eq:objective}, the optimization problem is conducted on dataset $\mathcal{D}$ with respect to parameters $\boldsymbol{\phi}, \boldsymbol{\theta}, \boldsymbol{\varphi}$ by addressing the optimization problem $\min_{\boldsymbol{\phi},\boldsymbol{\theta},\boldsymbol{\varphi}} \mathbb{L}(\mathcal{D}; \boldsymbol{\phi},\boldsymbol{\theta},\boldsymbol{\varphi})$ by means of stochastic gradient descent algorithm.

\begin{figure}[htbp]
    \begin{center}
        \subfigure[]{
        \begin{minipage}[t]{0.95\linewidth}
        \centerline{\includegraphics[width=0.9\textwidth]{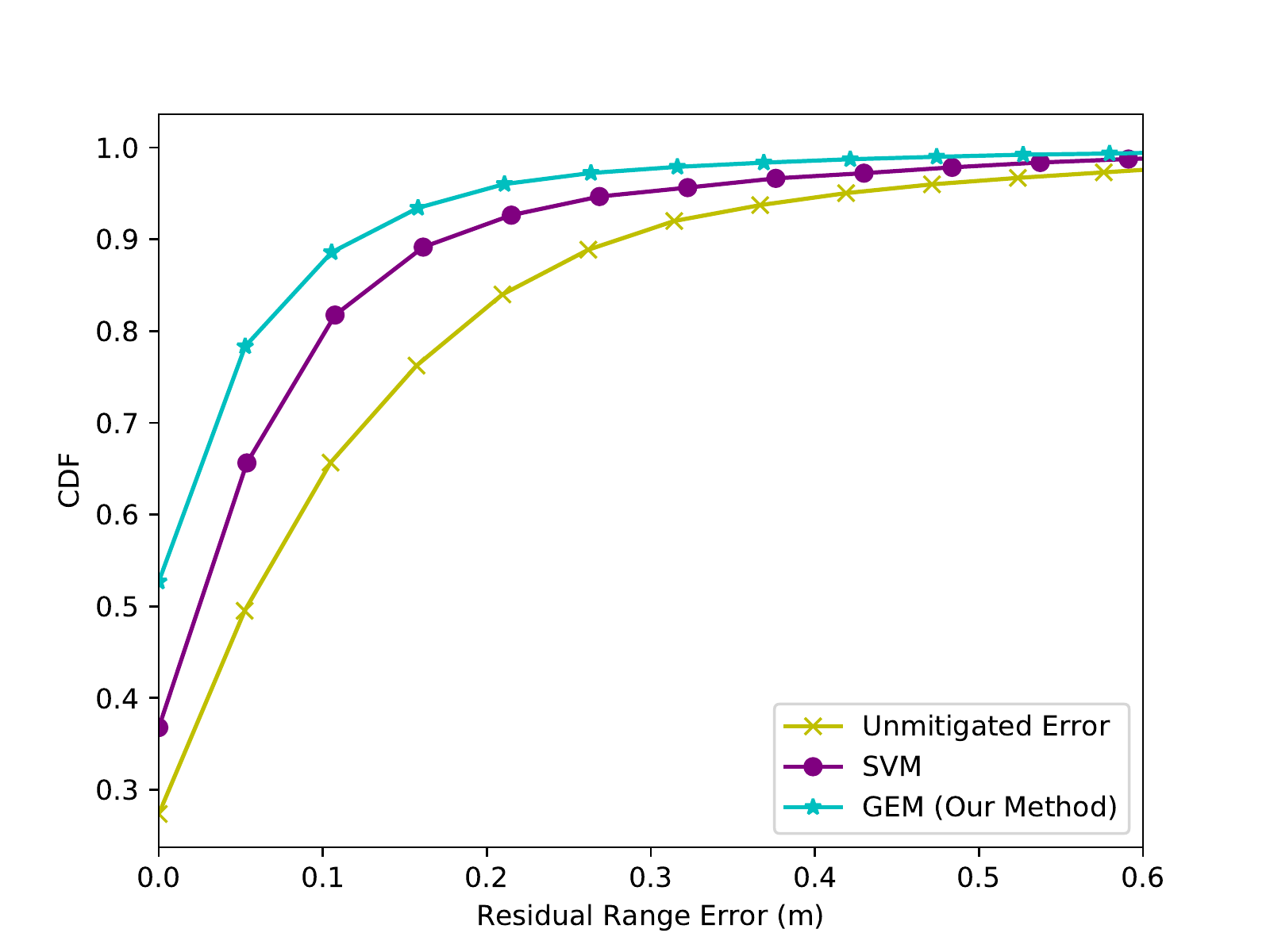}}
        \end{minipage}%
        }  \\
        \vskip -0.08in
        \subfigure[]{
        \begin{minipage}[t]{0.95\linewidth}
        \centerline{\includegraphics[width=0.9\textwidth]{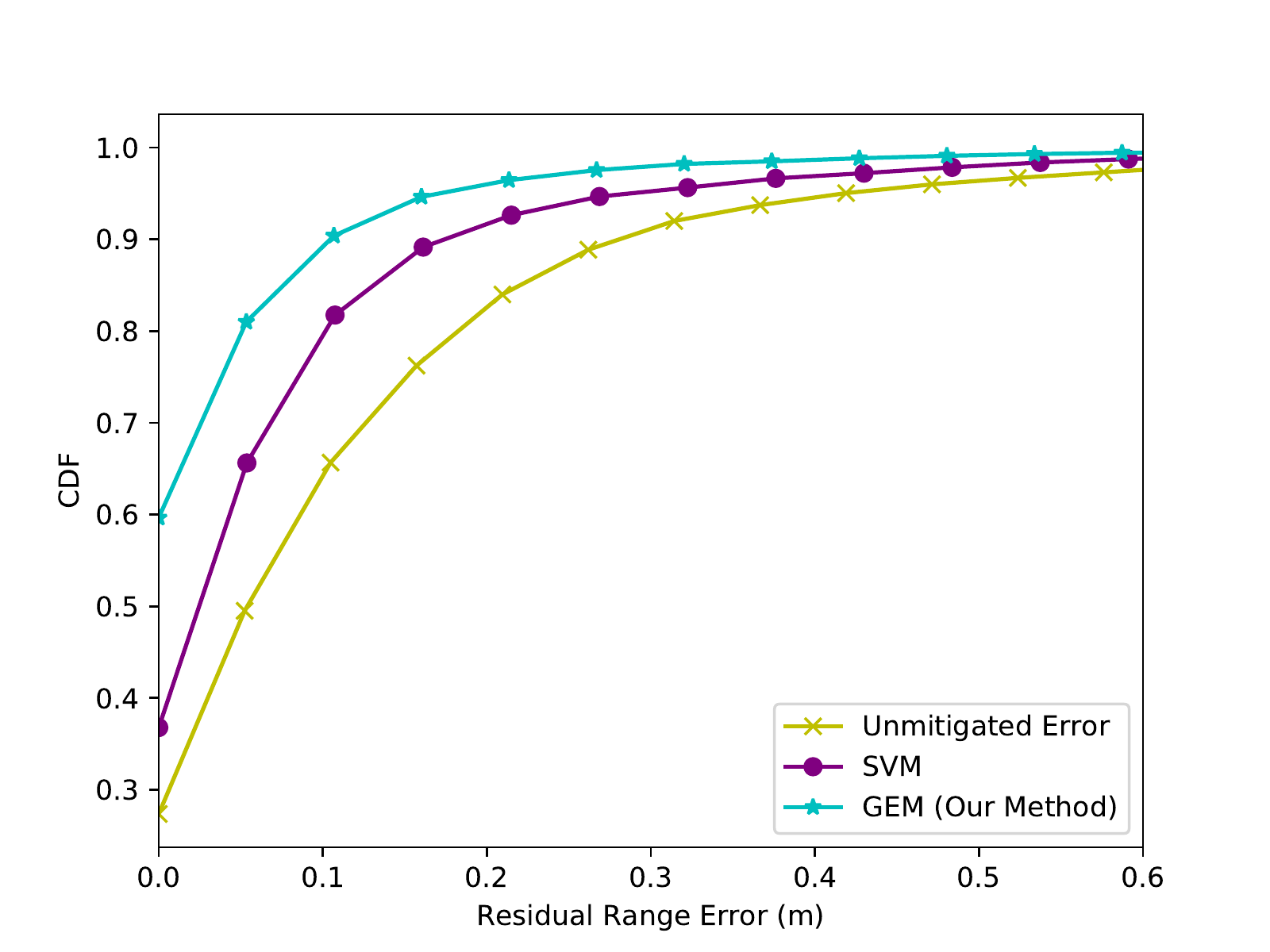}}
        \end{minipage}
        }
        \caption{The CDFs of the remaining error after mitigation on (a) \textit{Room Full} dataset, and (b) \textit{Obstacle Full} dataset.}
        \label{fig:CDFs}
    \end{center}
\end{figure}

\section{Experiments}
\label{sec:exp}



In this section, we evaluate the proposed algorithm using a UWB system as a case study. In particular, we give the quantitative results on range error mitigation and environment identification. Qualitative results of latent space visualization are also presented to show the effectiveness of environment semantic disentanglement.

\subsection{UWB Database and Experimental Setup}

\begin{figure*}[htbp]
    \begin{center}
        \includegraphics[width=0.9\textwidth]{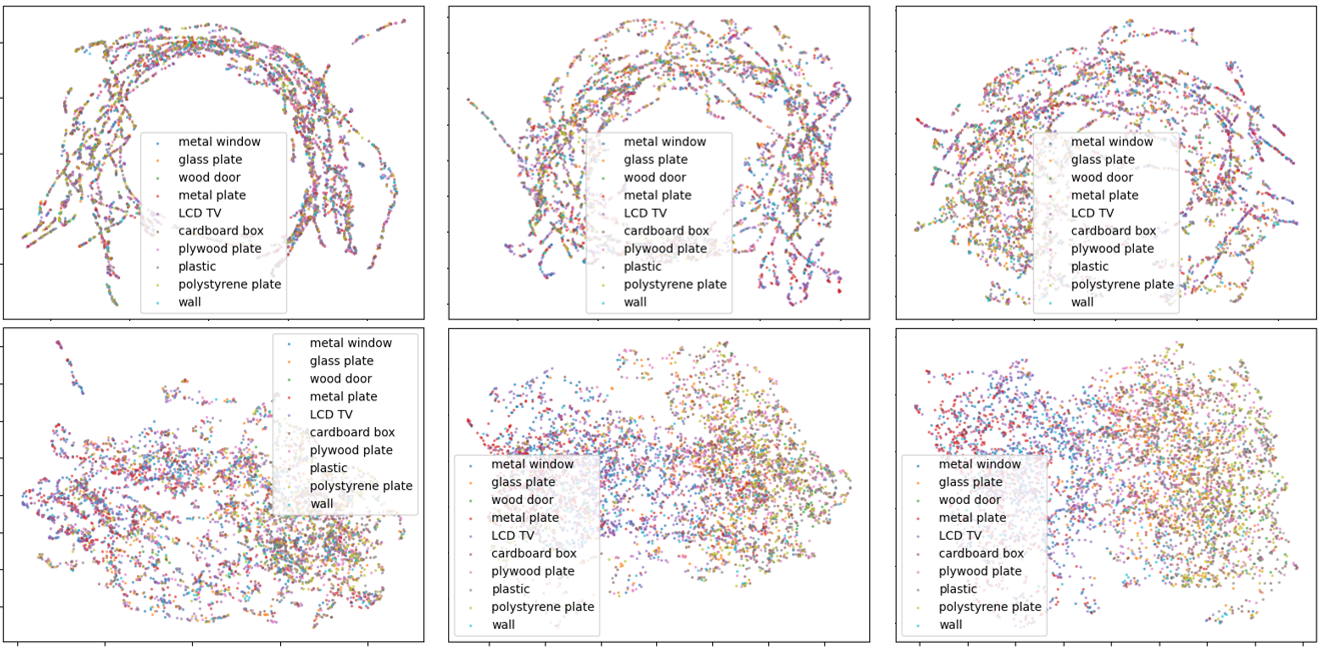}
        \vskip -0.1in
        \caption{Visualization of environment codes on the \textit{Obstacle Full} dataset, with model sampled every $100$ epochs from epoch $0$ to $500$.}
        \label{fig:latent_process}
    \end{center}
\end{figure*}

We use a public database \cite{zenodo} that is composed by waveform samples with labeled range errors (m) in different environment scenarios. The configurations for environment include $5$ room scenarios and $10$ obstacle scenarios. Specifically, the $5$ room scenarios include i) an outdoor space, ii) $3$ office-like rooms of large, medium, and small sizes, and iii) cross room measurements. The obstacle scenarios include $10$ different materials that block the LOS path. In order to explore the effect of different environments on waveforms, as well as evaluate the generality of the proposed method, we create $5$ different datasets from this database.

\begin{enumerate}
    \item \textit{Room Full}: waveform samples from all the $5$ room scenarios, each labeled with range error and the corresponding room scenario.
    \item \textit{Room Rough}: waveform samples from all the $5$ room scenarios, each labeled with range error and the rough 'indoor' or 'outdoor' settings (i.e., conclude $3$ office-like rooms and the cross room setting together as 'indoor').
    \item \textit{Room Part}: waveform samples from $3$ office-like rooms, each labeled with range error and the corresponding room scenario (i.e., big, medium or small sized).
    \item \textit{Obstacle Full}: waveform samples from all the $10$ obstacle scenarios, each labeled with range error and the corresponding obstacle scenarios.
    \item \textit{Obstacle Rough}: waveform samples with heavy blocking materials (i.e., 'metal') and light ones (i.e., 'plastic', 'wood', 'glass'), each labeled with range error and the rough 'heavy' or 'light' settings.
\end{enumerate}

For all the datasets, we randomly choose $80\%$ data samples as the training set and the rest $20\%$ samples as the testing set, without overlapping between the two sets to prevent overfitting.

We then conduct range error mitigation and environment identification on each of these datasets. 
SVM methods with hand-crafted features \cite{MazConAllWin:J18,WymMarGifWin:J12} are adopted as baselines. Since conventional methods cannot conduct the two tasks simultaneously, a SVM as in \cite{WymMarGifWin:J12} is trained as a regressor for range error, and a separate SVM as in \cite{XiaWenMarTriBluFro:J15} is trained as a classifier for environment identification. For convenience, we refer to the first SVM as SVR, and the second as SVC.

The proposed method is referred to as DGM for convenience. We build the VAE module and the classifier with cascaded $2$D convolutional layers, and the estimator module with linear layers. The whole model is trained on GTX $1080$ GPU with the accelerator powered by the NVIDA Pascal architecture. The code and trained models will all be released to public in our final version.

\subsection{Results of Range Error Mitigation}

We evaluate the range error mitigation performance in terms of the root mean square error (RMSE) and the mean absolute error (MAE). The results are presented in the first $6$ columns of Table \ref{tab:exp_comp}. The CDFs of the methods on dataset \textit{Room Full} and \textit{Obstacle Full} are shown in Fig.\ref{fig:CDFs}. It can be seen that the proposed DGM achieves superior results in all the datasets, implying both effectivess and generality. Specifically, DGM can realize a centimeter-level accuracy, with improvements to SVR of over above $55\%$ for RMSE and $80\%$ for MAE. The steady performance rise across dataset illustrates the generality of DGM, without a problem to certain dataset. In addition, DGM obtains better results on room-related datasets than on obstacle-related datasets. This implies that the obstacle materials have a more complicated impact on range error than room layouts.

\subsection{Results of Environment Identification}

We evaluate the identification performance in terms of classification accuracy of environment labels, shown in the last two columns of Table \ref{tab:exp_comp}.
The proposed DGM shows better results than SVC in all the datasets.
It can be noticed that the accuracy values on obstacle settings are relatively lower compared to room settings, in accordance with the results from range error mitigation that obstacles are more complex in semantics than room layouts.

\subsection{Latent Space Structure for Environment Semantics}

\begin{figure}[htbp]
    \begin{center}
        \includegraphics[width=0.45\textwidth]{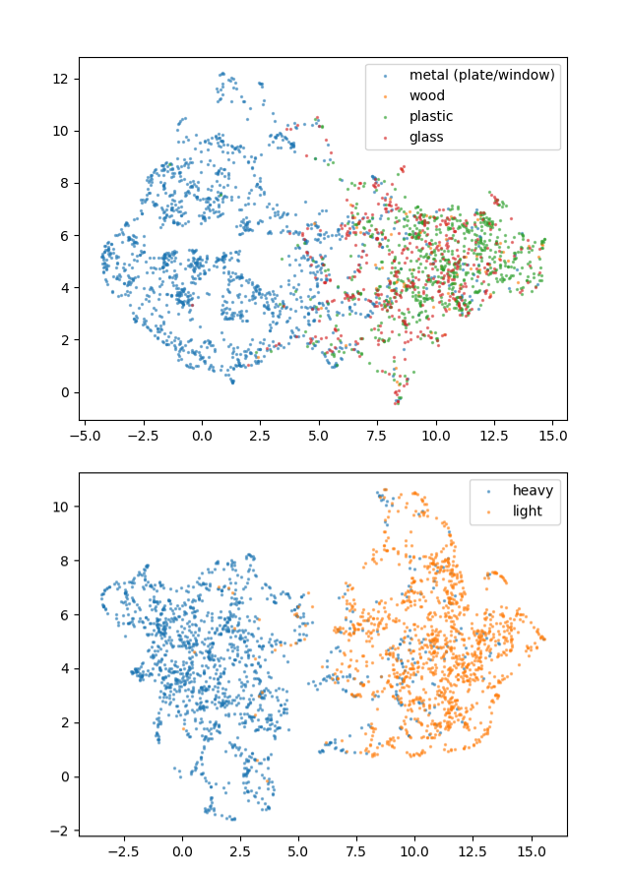}
        \caption{Visualization of environment codes in different obstacle scenarios.}
        \label{fig:latent_o}
    \end{center}
\end{figure}

\begin{figure}[htbp]
    \begin{center}
        \includegraphics[width=0.45\textwidth]{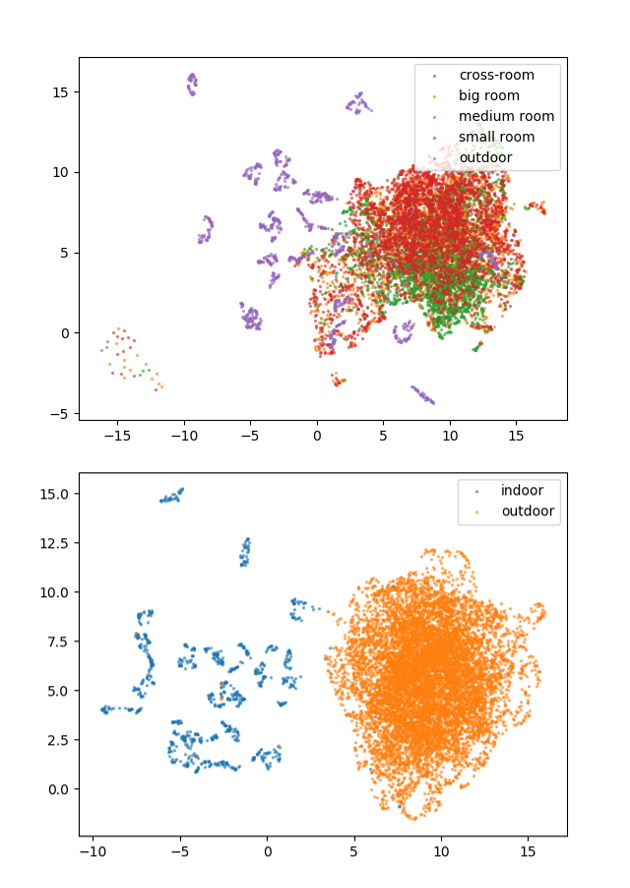}
        \caption{Visualization of environment codes in different room scenarios.}
        \label{fig:latent_r}
    \end{center}
\end{figure}

We use UMAP algorithm \cite{MciHea:J18} for latent space visualization of environment semantics, which can reduce the high-dimensional spaces to two dimensions while preserving the neighborhood of the latent codes. The scatter plots are 
shown in Figs.\ref{fig:latent_process}-\ref{fig:latent_r}. Each point is the latent code for one sample from the test set and the color represents the corresponding environment label. Note that the input to UMAP is the environment code $\mathbf{z}$, and the scales of the x- and y-axis don’t have any specific meaning but pure illustration for 2-dimensional visualization.

The learning process of latent codes is shown in Fig.\ref{fig:latent_process}. It can be seen that code samples are gathered in a ring-shaped manifold in the early epochs. With the learning progress, the ring is gradually unfolded to fulfill the $2$-dimensional space. Moreover, a spectrum can be observed, with heavy materials (e.g. metal windows (blue), metal plate (red), LCD TV (purple)) presented in the left part and light materials (e.g. glass plate (orange), wood door (green)) in the right part. 
We further consider latent codes of several critical classes for clearness, 
illustrated in Fig.\ref{fig:latent_o}. It can be seen that metal obstacles (blue) is well separated from the others, while the rest three materials (red-green-orange) form a cluster. This hints at a connection between these environments, as they are all relatively light materials of low density. In the latter case, a remarkable separation of the two class can be observed. This implies that objects are clustered based on different dielectric coefficients, which consistent with the intuition.

Similar visualizations are conducted for room scenarios, shown in Fig.\ref{fig:latent_r}. 
We can observe a cluster of indoor scenarios (red-green-orange), and a separation of the indoor and outdoor scenarios (purple).
This explains the drastic difference between indoor and outdoor environments, and why existing methods find it challenging to generalize in both indoor and outdoor conditions. 
Such phenomenon implies that the environmental semantics lies in a low-dimensional manifold embedded in the high-dimensional data space, and can be effectively disentangled by the proposed method.

\section{Conclusion}
\label{sec:con}

The paper introduced a DGM for efficient feature extraction of waveforms, which can simultaneously conduct range error mitigation and beyond NLOS environment identification. The proposed method was based on a Bayesian model and implemented by an efficient end-to-end learning network. The complicated high semantic features in raw waveform data were automatically exploited via the presented DGM framework. Experimental results illustrated the superior performance of our method on both tasks using a general dataset with different environmental settings. The presented methodology also guaranteed potential variants to the extraction of soft range information and direct learning-based localization schemes, which will be our future work.

\section*{Acknowledgment}

This research is partially supported by the Basic Research Strengthening Program of China (173 Program) (2020-JCJQ-ZD-015-01), the  Basque Government through the ELKARTEK programme, the Spanish Ministry of Science and Innovation through Ramon y Cajal Grant RYC-2016-19383 and Project PID2019-105058GA-I00, and Tsinghua University - OPPO Joint Institute for Mobile Sensing Technology.

\bibliographystyle{IEEEtran}
\bibliography{IEEEabrv,StringDefinitions,SGroupDefinition,refs}

\end{document}